\documentclass[%
 reprint,
 amsmath,amssymb,
 aps,
]{revtex4-2}

\usepackage{xcolor}
\usepackage{graphicx}% Include figure files
\usepackage{dcolumn}% Align table columns on decimal point
\usepackage{bm}% bold math
\usepackage{hyperref}% add hypertext capabilities
\usepackage[normalem]{ulem}
\usepackage{gensymb}
\usepackage{wasysym}
\usepackage[normalem]{ulem}
\begin{document}

\preprint{APS/123-QED}

\title{Searching for alignment-to-orientation conversion in the ground state of atomic Cs with circularly polarized laser probe}

\author{A. Mozers}
 \email{arturs.mozers@lu.lv}
\author{L. Busaite}
\author{D. Osite}
\author{F. Gahbauer}
\author{M. Auzinsh}
\affiliation{
 Laser Centre, University of Latvia, Rainis Boulevard 19, LV-1586 Riga, Latvia
}

\date{\today}% It is always \today, today, but any date may be explicitly specified

\begin{abstract}
In this study we explored the possibilities for observing the angular momentum alignment-to-orientation  conversion (AOC) in the ground state of various alkali metals: K, Rb, Cs. For theoretical analysis we used a model that is based on the Optical Bloch equations for the density matrix. Our model includes the interaction of all neighboring hyperfine levels with laser radiation, the mixing of magnetic sublevels in an external magnetic field, the coherence properties of the exciting laser radiation, and the Doppler effect. Additionally we simulated signals where the ground- or the excited-state coherent processes were numerically switched off in order to determine the origins of the features of the obtained signals. We also performed experiments on Cs atoms with two laser beams: a linearly polarized Cs $D_1$ pump and circularly polarized Cs $D_2$ probe. We used the pump beam to create angular momentum alignment in the ground state and observed the transmission signal of the probe beam as we changed the magnetic field. A detailed analysis of the experimentally obtained transmission signal from a single circularly polarized probe laser component is provided. Finally, prospects for observing AOC conversion experimentally are discussed, as well as experiments were even a weak AOC signal could lead to systematic errors.

%\begin{description}
%\end{description}
\end{abstract}

\maketitle

%\tableofcontents

\section{\label{sec:level1}Introduction\protect}

Symmetry appears not only at the heart of fundamental physical concepts, such as CPT symmetry, but also in more mundane phenomena such as absorption of light by an ensemble of atoms. These two levels are connected in sensitive experiments that search for evidence of symmetry-breaking on the fundamental level by detecting, for example, circularly polarized light where only linearly polarized light would be expected.
When linearly polarized light interacts with an ensemble of atoms, the angular momentum distribution of the excited atoms becomes aligned with the polarization vector of the exciting radiation. An aligned state is created with inversion symmetry about the axis of alignment. In general, the aligned excited state will emit linearly polarized fluorescence, and, in doing so, it may generate alignment in the ground state as well. Now, if there is an additional interaction that breaks the inversion symmetry, the aligned state may be transformed into an oriented state, which lacks inversion symmetry about the axis of orientation and may emit circularly polarized light. This effect is called alignment-to-orientation conversion (AOC). AOC has been achieved with magnetic field gradients~\cite{Fano:1964} and anisotropic collisions~\cite{Lombardi:1967, Rebane:1968, Manabe:1981}. Another common way to realise AOC is to use an external electric field~\cite{Lombardi:1969,Auzinsh:2006}. In contrast to the electric field, an external magnetic field by itself cannot cause AOC, because it is an axial field and has even parity. However, when the interaction of the atoms with the magnetic field is comparable to the hyperfine interaction within the atom, the Zeeman effect becomes nonlinear and the axial symmetry is lost. AOC becomes possible~\cite{Alnis:2001,Lehmann:1964, Baylis:1968, Lehmann:1969, Krainska-Miszczak:1979,Mozers:2015}.

\begin{figure}[t]
	\includegraphics[width=0.7\linewidth]{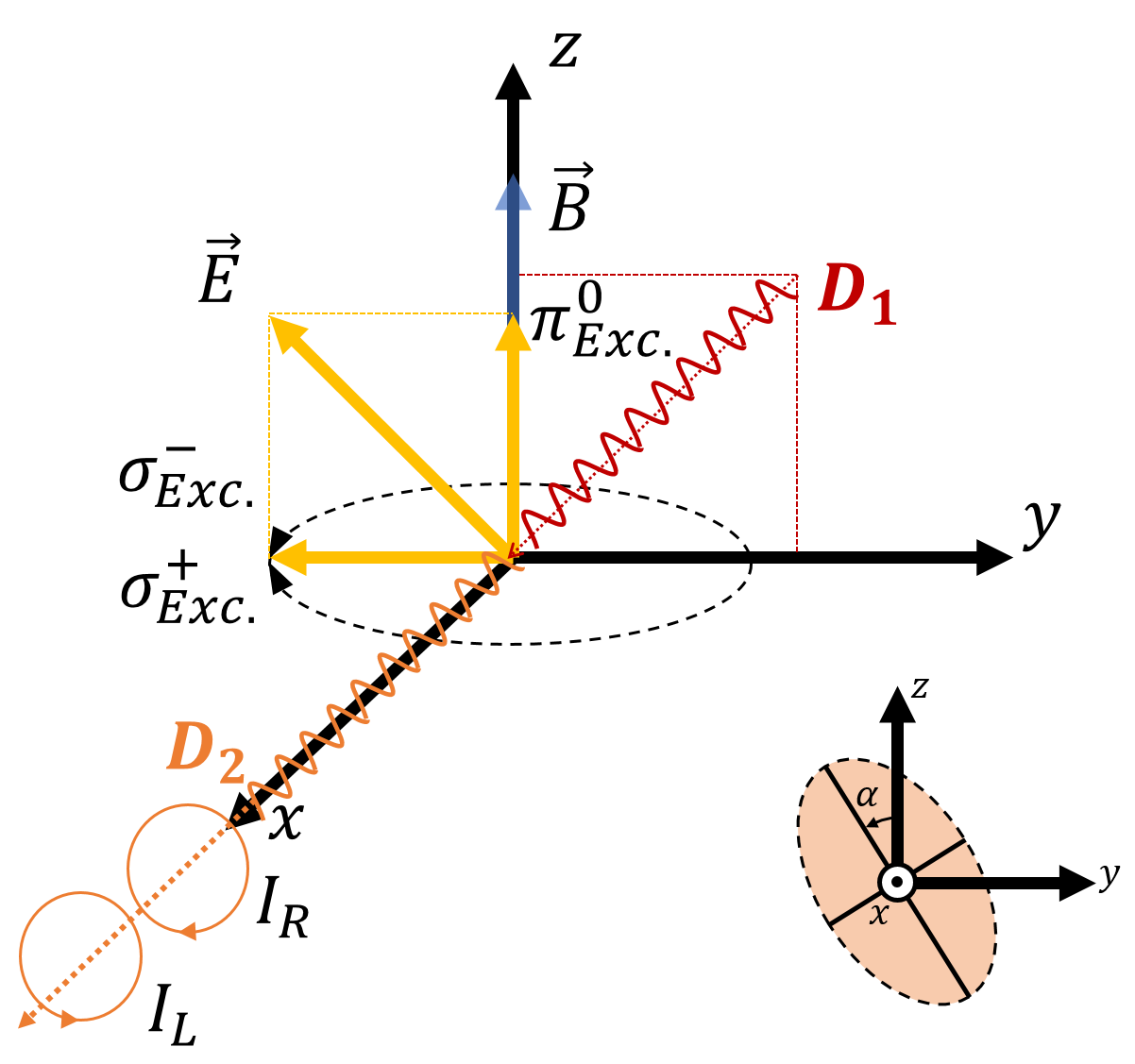}
    \caption{\label{fig:geom}Excitation and probing geometry. Linearly polarized Cs $D_1$ pump beam and circularly polarized Cs $D_2$ probe beam. The inset shows the azimuth angle $\alpha$ of the elliptically polarized probe light.}
\end{figure}

\begin{figure}[h!]
	\includegraphics[width=\linewidth]{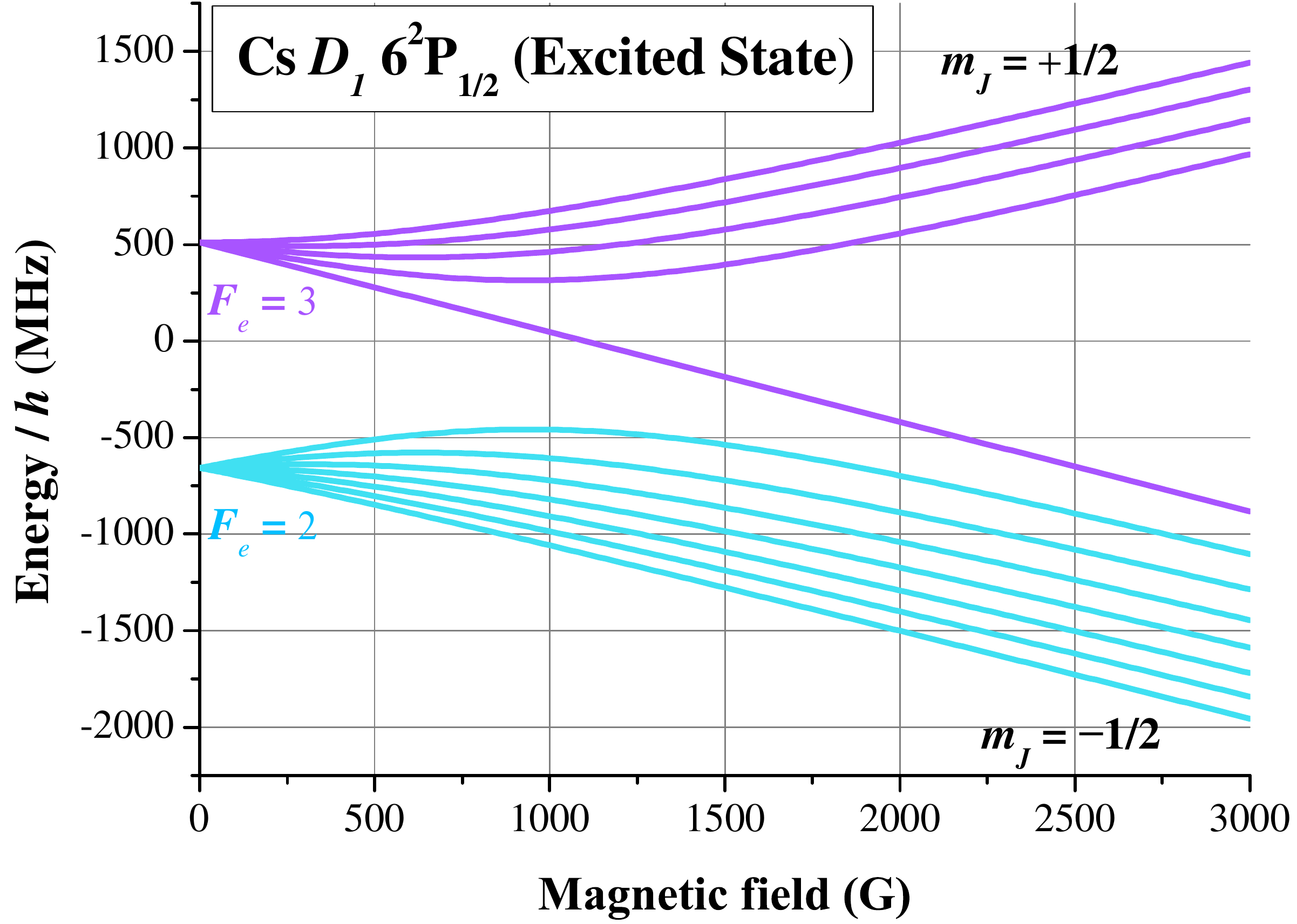}
    \caption{\label{fig:Breit-Rabi-ex}Frequency shifts of the magnetic sublevels $m_F$ of the excited-state fine-structure level 6$^2$P$_{1/2}$ as a function of magnetic field for $^{133}$Cs. Zero frequency shift corresponds the excited-state fine-structure level 6$^2$P$_{1/2}$.}
\end{figure}

\begin{figure}[h!]
	\includegraphics[width=\linewidth]{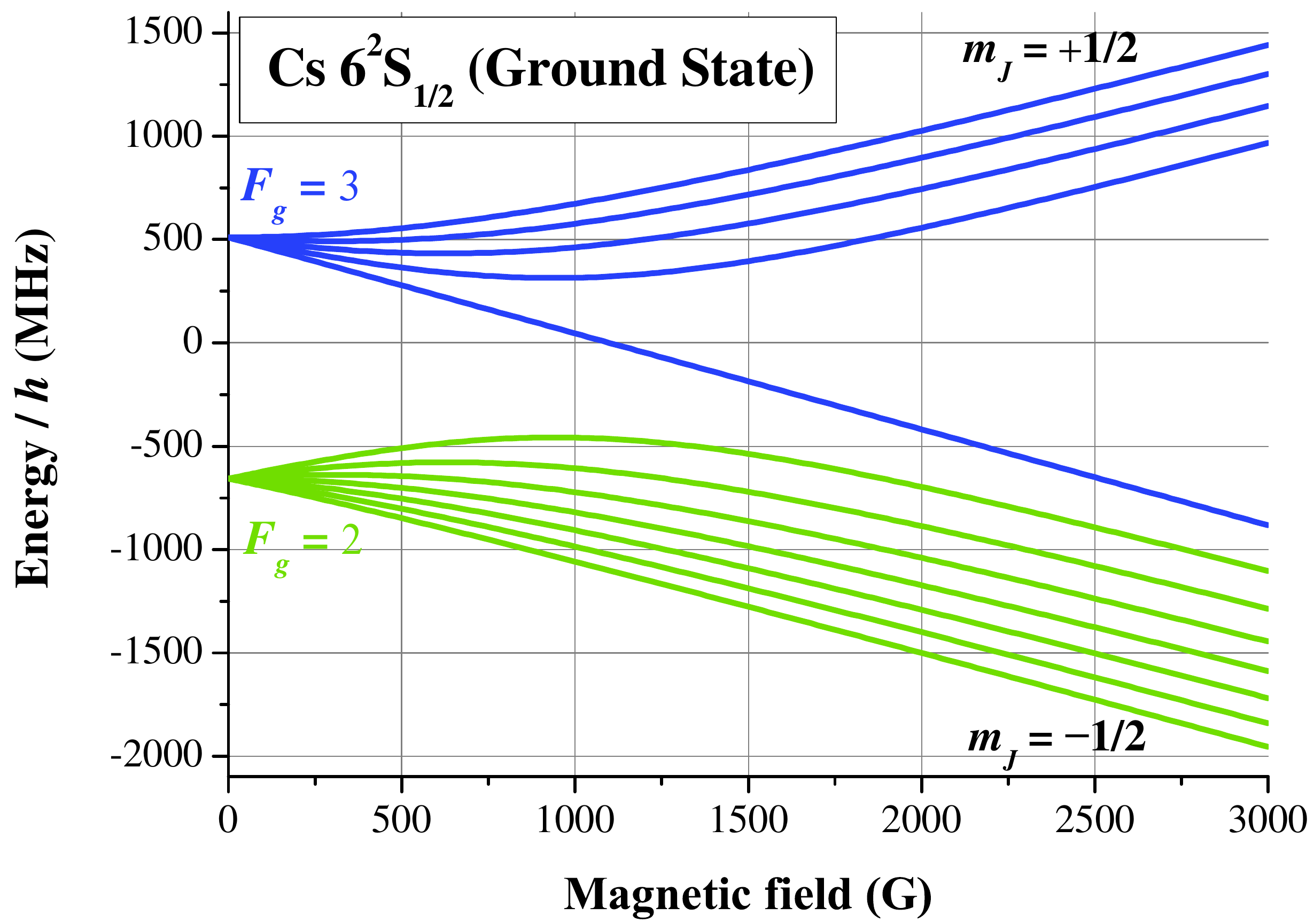}
    \caption{\label{fig:Breit-Rabi-gr}Frequency shifts of the magnetic sublevels $m_F$ of the ground-state fine-structure level 6$^2$S$_{1/2}$ as a function of magnetic field for $^{133}$Cs. Zero frequency shift corresponds to the ground-state fine-structure level 6$^2$S$_{1/2}$.}
\end{figure}

Alignment is longitudinal if it coincides with the quantization axis and transverse if it is perpendicular to the quantization axis. In the latter case, alignment is characterized by coherences among magnetic sublevels whose magnetic quantum numbers differ by $\Delta m_F=\pm 2$. Similarly, orientation is longitudinal when it is parallel to the quantization axis and transverse when it is perpendicular to the quantization axis. In the latter case, there are coherences among magnetic sublevels with $\Delta m_F=\pm 1$. 
In this work we are interested in the conversion of transverse alignment to transverse orientation. An initially aligned state can be transformed into transverse orientation by, first, setting the linearly polarized excitation at an angle of $\pi/4$ with respect to the quantization axis (Fig.~\ref{fig:geom}) so that the light polarization can be expressed as a superposition of one linear and two circular polarization components. Second, it is necessary to apply an external magnetic field so that the conditions for the hyperfine Paschen-Back effect are fulfilled and the dependence of the energies of the magnetic sublevels is nonlinear (Figs.~\ref{fig:Breit-Rabi-ex},~\ref{fig:Breit-Rabi-gr}). When all three optical components of light polarization are present the coherent excitation of $\Delta m_F=\pm 1$ is possible, which leads to the creation of transverse orientation of angular momentum \cite{Auzinsh:2010book}.

The ground-state AOC merits further investigation because the generation of orientation can be a sensitive background in experiments that rely on the symmetry of aligned angular momentum. However, precisely because it is a potentially tiny background, the effect is difficult to measure. We had studied ground-state AOC in Rubidium previously by observing fluorescence signals~\cite{Mozers:2020}. Since fluorescence directly involves the excited state, it was challenging to disentangle excited-state AOC from ground-state AOC. Intuitively, measuring the ground-state absorption signal should lead to results that are easier to interpret. Nevertheless, the signals are very small relative to background. Therefore, we aim to explore the conditions that could generate measurable AOC using theoretical calculations backed up by experimental measurements that validate the theoretical approach. The results of signal simulation would guide an experiment used to measure ground-state AOC directly or be useful when estimating the potential influence of the AOC effect on experiments that search, for example, for the permanent electric dipole moment of an electron~\cite{Cairncross2019} or parity violation (PV) in atomic physics~\cite{Roberts:2015}.

%\subsection{\label{sec:level2}Second-level heading: Formatting}
%\subsection{\label{sec:citeref}Citations and References}
%\subsubsection{Citations}
%\paragraph{Syntax}
%\paragraph{Eliding repeated information}
%\subsubsection{Example citations}
%\subsubsection{References}
%\subsubsection{Example references}
%\subsection{Footnotes}%

\section{Theoretical Model} \label{theory}

The theoretical model, based on the Liouville or optical Bloch equations (OBEs) for the density matrix $\rho$, is used to describe absorption of circularly and elliptically polarized light in an atomic vapor subjected to a strong pump laser field and external magnetic field. The density matrix is written in the basis of the ground- and excited-state hyperfine levels $\vert\xi, F_i, m_{F_i}\rangle$, where $F_i$ refers to the total angular momentum quantum number of the hyperfine level, $i$ refers to the ground $g$ or excited state $e$, $m_{F_i}$ denotes the respective magnetic quantum number and $\xi$ represents all the other quantum numbers that are not essential for the current study.

In the first step we describe the time evolution of the density matrix of an atom in a strong laser field and magnetic field using the optical Bloch equations~\cite{Stenholm:2005}
\begin{equation}\label{eq:liouville}
	i\hbar\frac{\partial \rho}{\partial t} = \left[\hat H,\rho\right] + i\hbar\hat R\rho,
\end{equation}

where $\hat H$ is the total Hamiltonian operator of the system and $\hat R$ is the relaxation operator. The full Hamiltonian can be written as $\hat H = \hat H_0 + \hat H_B + \hat V$, where $\hat H_0$ is a Hamiltonian of the free atom, $\hat H_B$ describes the interaction with the external magnetic field, and $\hat V = -\hat{\mathbf{d}}\cdot \mathbf{E}(t)$ is the operator that describes the atom-laser field interaction in the electric dipole approximation. The operator includes the electric field of the exciting light $\mathbf{E}(t)$ and an electric dipole operator $\hat{\mathbf{d}}$.

As we use continuous laser excitation, we calculate the density matrix for steady state conditions 
\begin{equation}\label{eq:steady}
	\frac{\partial \rho_{g_ig_j}}{\partial t} = \frac{\partial \rho_{e_ie_j}}{\partial t} = 0,
\end{equation}
where $\rho_{g_ig_j}$ and $\rho_{e_ie_j}$ are density matrix elements for the ground and exited state, respectively.
This condition reduces the differential equations to a system of linear equations. The explicit form of these equations, the description of terms appearing in these equations and solution methods can be found in~\cite{Blushs:2004, Mozers:2020}. 

For the analysis of the results, the multipole expansion of the density matrix can be used. Expansion coefficients can be calculated as~\cite{Auzinsh:2010book}:
\begin{equation}\label{eq:m_moments}
	\rho_{q}^{\kappa} = \sum_{mm'}(-1)^{F-m'}\langle FmF,-m'\vert \kappa q\rangle\rho_{m'm},
\end{equation}
where $\rho_q^\kappa$ is the co-variant component of multipole moments, $\kappa$ is the rank of the polarization moment with $2\kappa + 1$ components $q = -\kappa, \dots \kappa$.
From Eq.~\eqref{eq:m_moments} we can determine that the angular-momentum transverse orientation $\rho_{\pm1}^{1}$ is constructed from the first off-diagonal elements in the calculated density matrix, which correspond to $\Delta m_F=\pm 1$ coherences.

To compare the model calculations with experimentally measured signals we need to average the calculated absorption values over the Doppler profile.

To probe the ground state, we can calculate absorption of left- and right-circularly polarized laser radiation tuned to the $D_2$ line.
It is assumed that the probe beam is much weaker than the pump beam, so that the former does not disturb the state prepared by the latter.

The absorption of the weak probe beam of polarization $\mathbf{e}_\text{probe}$ is calculated as \cite{Auzinsh:2009b}
\begin{equation}
    A(\mathbf{e}_\text{probe}) = \tilde{A}_0 \sum_{e_ig_jg_k} \frac{d_{e_ig_j}^\text{(probe)} \rho_{g_jg_k} d^{\ast\text{(probe)}}_{g_k e_i} }{\Delta_{e_ig_j}^2 + \Gamma_R^2},
\end{equation}
where $d_{e_kg_i}^\text{(probe)} = \langle {e_k} | {\bf\hat d} \cdot {\mathbf{e}_\text{probe}} | {g_i} \rangle$ are the dipole transition matrix elements for the radiation with specific polarization ($\mathbf{e}_\text{probe}$).
$\tilde{A}_0$ is a constant of proportionality.

The detuning from resonance $\Delta_{e_ig_j}$ is expressed as:
\begin{equation}
    \Delta_{e_ig_j} = \bar\omega - \omega_{e_ig_j}
\end{equation}
$\bar\omega$ is the central laser frequency and $\omega_{e_ig_j}$ is energy difference between magnetic sublevels $\vert e_i\rangle$ and $\vert g_j\rangle$.

The absorption width is defined by relaxation rate $\Gamma_R$
\begin{equation}
    \Gamma_R = \frac{\gamma + \Gamma + \Delta\omega}{2},
\end{equation}
which is influenced by transit relaxation rate $\gamma$, the rate of spontaneous transitions $\Gamma$ and spectral width $\Delta\omega$ of laser radiation.

The average transit relaxation rate $\gamma$ can be estimated as 
\begin{equation}\label{eq:gamma}
	\gamma = \frac{v_{th}}{d},
\end{equation}
where $v_{th}$ is the thermal velocity of atoms in the plane perpendicular to the laser beam and $d$ is the laser beam diameter.
In the calculation the transit relaxation was assumed to be the same for both processes: pumping and probing. 

The reduced Rabi frequency of the pump beam was estimated as \cite{Auzinsh:2016}
\begin{equation}\label{eq:Rabi}
	\Omega_R = \frac{\vert\vert d\vert\vert\cdot\vert\varepsilon\vert}{\hbar} = \frac{\vert\vert d\vert\vert}{\hbar}\sqrt{\frac{2I}{\epsilon_0 n c}},
\end{equation}

where $\vert\vert d\vert\vert$ is the reduced dipole matrix element for the $D_1$ transition~\cite{Auzinsh:2010book, Auzinsh:2009b}, $e$ is the electron charge, and $a_0$ is the Bohr radius, 
The quantity $I$ is the power density (proportional to the amplitude squared of the electric field $\vert\varepsilon\vert^2$), $\epsilon_0$ is the electric constant, $n=1$ is the refractive index of free space, and $c$ is the speed of light.

\section{Experiment}
\begin{figure}[h!]
	\includegraphics[width=\linewidth]{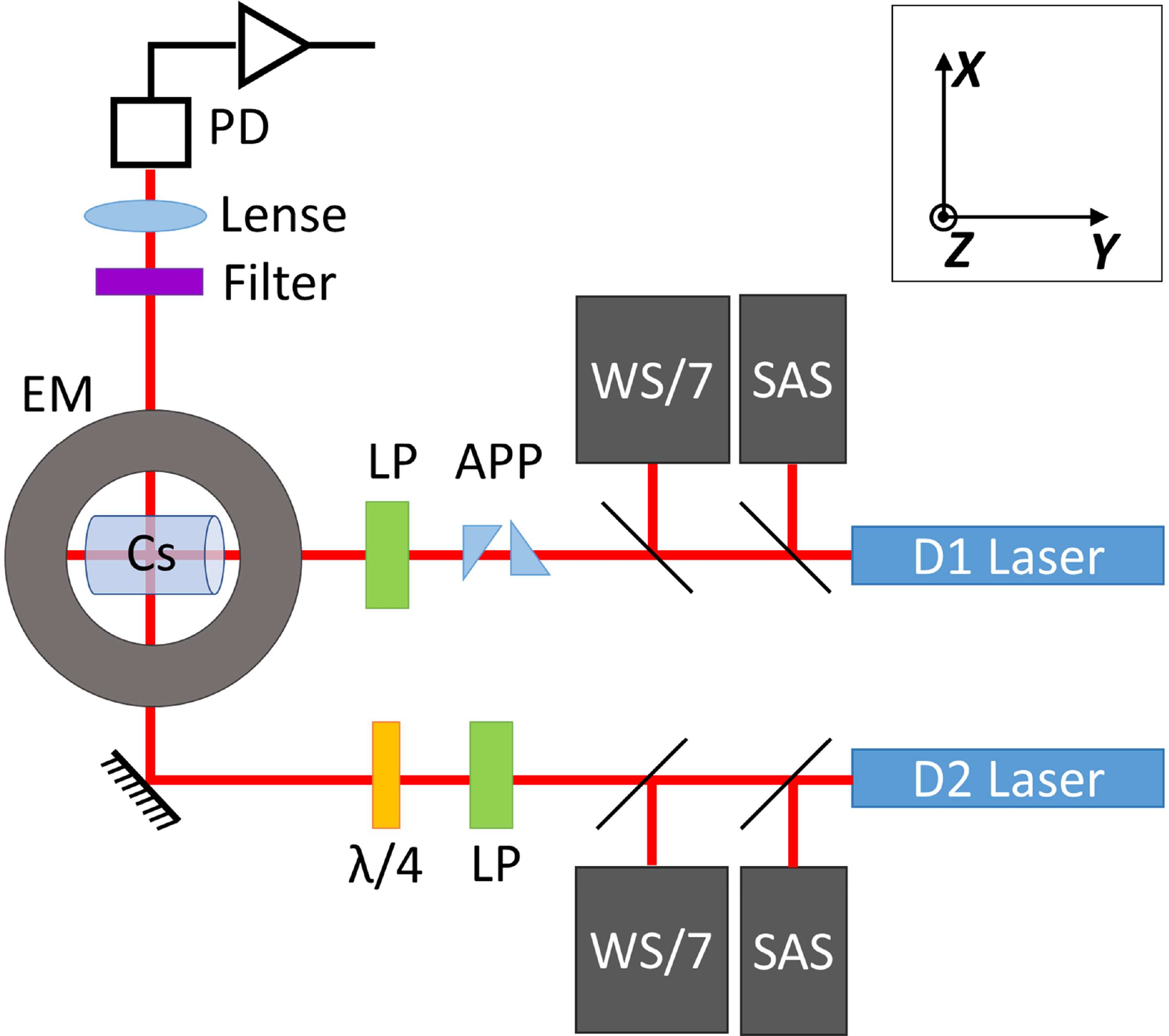}
    \caption{\label{fig:exp_setup}Side view of the experimental setup. The Cs $D_1$ and $D_2$ lasers are the pump and probe lasers, respectively. SAS is the saturation absorption spectroscopy setup. WS/7 is a wavemeter. APP is the achromatic prism pair. LP is a linear polarizer. $\lambda/4$ is an achromatic quarter wave plate. Cs is the atomic vapour cell inside the electromagnet (EM). PD is the photodiode.}
\end{figure}
To create the pump-probe geometry~(Fig.\ref{fig:geom}), we built an experiment which is shown schematically in Fig.~\ref{fig:exp_setup}. We placed a $\diameter$ 2.5 $\times$ 2.5 cm cylindrical pyrex Cs vapor cell between the ferromagnetic cores of an electromagnet (EM). The spacing between the poles was 4.3 cm.
The inhomogeneity of the field in the center of the poles was estimated to be no more than 0.027 \% by a finite-element electromagnetic field solver, based on the geometry of the magnet. The current for the electromagnet was supplied by a bipolar power supply, and the symmetrical triangular current scan was generated by a function generator. The frequency of the magnetic field scan was 2.00~mHz with a maximum scan amplitude resulting in a magnetic field range from $-3100$ to $+3100$~G.

The atoms in the cesium vapor cell were pumped to the excited state $6P_{1/2}$ by linearly polarized light from a DPSS Ti:Sapphire laser (SolsTiS from M Squared) with a wavelength of 895 nm ($D_1$ line of Cs) and probed with circularly polarized light from a tuneable, distributed feedback, single-mode diode laser DL 100 (Toptica AG) with a wavelength of 852 nm ($D_2$ line of Cs). 

In the interest of increasing the volume of pumped atoms along the observation direction the beam of the $D_1$ laser was expanded in the $x$ direction by a pair of anamorphic prisms (APP). The elliptically shaped beam dimensions were measured by a beam profiler. The semi-major axis was 5 mm, and the semi-minor axis was 2 mm as measured by a Gaussian fit. The pump laser beam was directed to the cesium vapor cell so it would enter the cell at a $\pi/4$ angle with respect to the magnetic field direction.

The $D_2$ laser frequency was fixed to a saturation absorption spectrum (SAS) signal.
A dedicated feedback controller (DigiLock by Toptica AG) managed the laser's temperature and current controllers to lock the laser to a particular peak of the SAS. During the experiments the frequency of the laser was monitered by a HighFinesse WS/7 Wavemeter. The probe laser power was attenuated using a half-wave Fresnel rhomb retarder followed by a linear polarizer and neutral density filters. An achromatic quarter wave plate ($\lambda/4$) was used to convert the linear polarization of the probe beam into circular polarization.

The transmission signal was detected with a Thorlabs SM1PD1A photodiode (PD). The photodiode was placed at the end of a tube which also contained a convex lens and a shortpass filter.
The signal from the photodiode was amplified by a voltage amplifier with a gain of $10^7$. Every scan of each probe component was acquired with an Agilent DSO5014A digital oscilloscope and transferred to a personal computer.

The measured transmission signals were averaged over multiple scans. When comparing the experimental signals to signals obtained from the theoretical calculations, a constant background was subtracted before the signals were normalized to the maximum of each component.

\section{Results \& Discussion}

\begin{figure*}
    \includegraphics[width=0.99\textwidth]{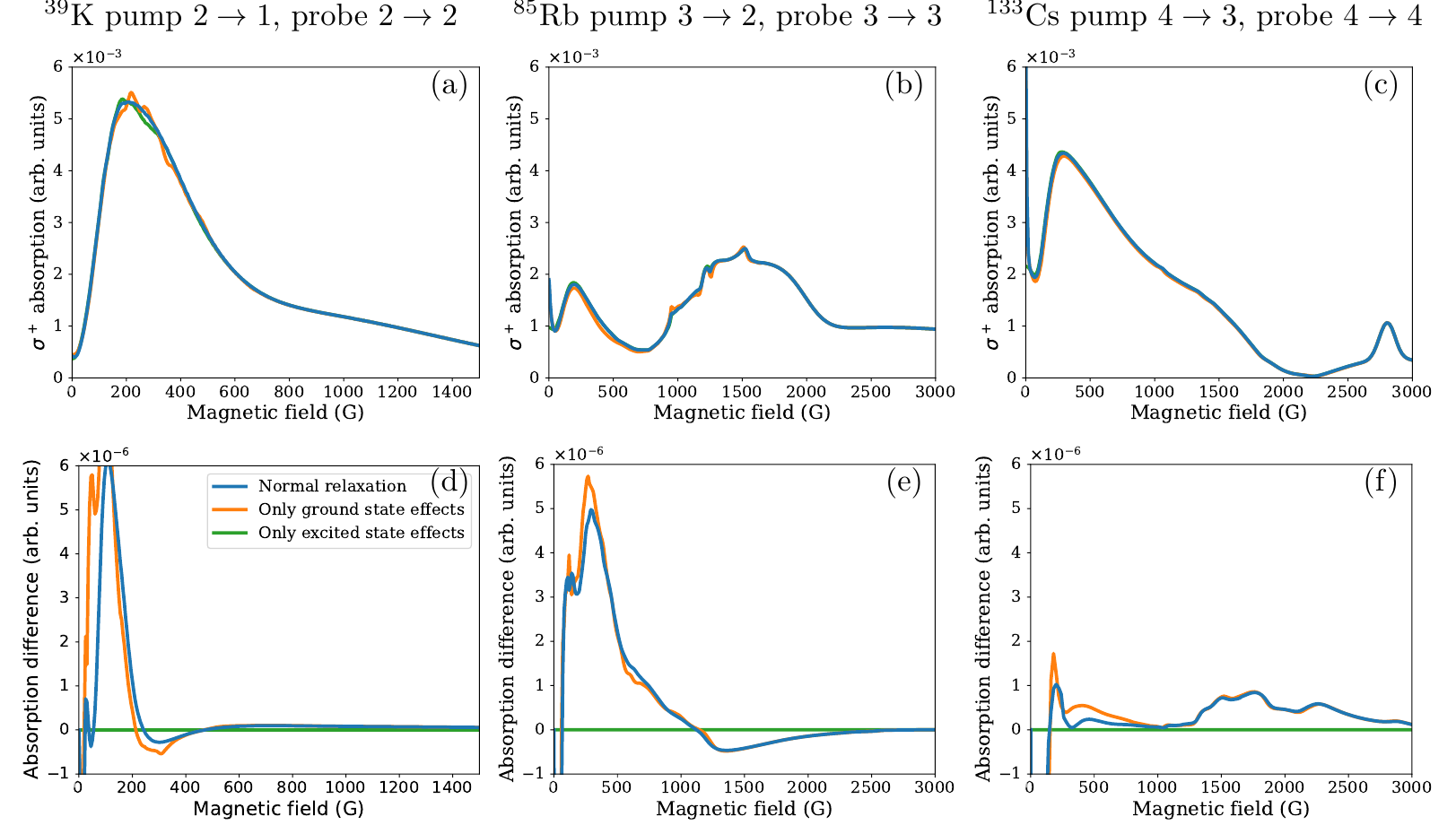}
    \caption{Numerical calculations of absorption for $^{39}$K, $^{85}$Rb and $^{133}$Cs. First row: Circular component of absorption. Second row: the difference of circular ($\sigma^+$ and $\sigma^-$) components. The pump and probe beam laser frequencies were fixed to the hyperfine transitions indicated at the top of each column. In all cases the Rabi frequency was $\Omega_R = 100~\mathrm{MHz}$.}
    \label{fig:abs_elements}
\end{figure*}

We wanted to test if the ground-state AOC would be easier to observe in absorption signals as compared to fluorescence signals, since fluorescence signals are more closely associated with effects that occur in the excited state. AOC in the ground state would be created when a linearly polarized pump beam interacts with the atoms in the presence of a magnetic field. The AOC would be detected by measuring the difference in the absorption of $\sigma^+$ and $\sigma^-$ circularly polarized probe beams. We call these signals absorption difference signals. We first modelled the absorption and absorption difference signals of circularly polarized probe beams for various alkali metal transitions in order to determine where the ground-state AOC might be most easily detected in the absorption (subsec.~\ref{sub_A}). Next, we obtained experimental signals of the absorption of circularly polarized light and analyzed the physical effects that influence the shape of these signals (subsec.~\ref{sub_B}). Finally, realizing that the absorption difference signals depend sensitively on the degree of circular polarization of the probe, we modeled such absorption difference signals for various non-ideal polarization configurations of the probe beam to determine how feasible the detection of ground-state AOC in these signals would be (subsec.~\ref{sub_C}).

\subsection{Simulated Absorption signals in different alkali atoms} \label{sub_A}

Figure~\ref{fig:abs_elements} shows the modeled results of absorption signals from the alkali-metal elements K, Rb and Cs. The left section (Fig.~\ref{fig:abs_elements} (a), (d)) shows the absorption signals from $^{39}$K atoms, the middle section (Fig.~\ref{fig:abs_elements} (b), (e)), from $^{85}$Rb, and the right section (Fig.~\ref{fig:abs_elements} (c), (f)), from $^{133}$Cs atoms. Indicated at the top of Figure~\ref{fig:abs_elements} are the pairs of hyperfine transitions for the pump and probe beams of the modeled signals that exhibited the highest values in the absorption difference signals and made it easier to set apart the features connected to ground-state coherences from the features connected to excited-state coherences.

In order to distinguish between the features in the signal caused by ground-state and excited-state coherent effects, we performed the following numerical experiment.
We switched off coherences either in the excited or in the ground state when carrying out the calculations using the theoretical model from Sec.~\ref{theory}. Technically we realized this by setting the relaxation rate of the non-diagonal elements in the excited state or in the ground state to a very high value. Namely, we put $\gamma_{non-diagonal}/\gamma_{diagonal} = 10^9$ either in the excited state or in the ground state. The relaxation rate $\gamma_{diagonal}$ was left unchanged for all cases and its value assumed equal to the transit relaxation rate that we expect in the experiment (see~\ref{eq:gamma}). This allows us to observe the influence of the transfer of coherences from the ground state to the excited state (and vice versa). Each subfigure of Fig.~\ref{fig:abs_elements} shows the comparison of the three different cases of simulated absorption difference signals:
the blue curve corresponds to the case when all elements in the density matrix experience normal relaxation---both ground- and excited-state coherent effects appear in the signal; the orange curve corresponds to the case when the excited-state coherent effects were set to zero---only ground-state coherent effects are present in the absorption signal; the green curve corresponds to the case when the ground-state coherent effects were set to zero---only excited-state coherent effects appear in the signal. These alterations to the $\gamma_{non-diagonal}$ have only a tiny effect on the absorption signal of a single circularly polarized component, the change never exceeding a few percent (figures~\ref{fig:abs_elements} (a), (b), (c)).

Figure~\ref{fig:abs_elements} (d), (e) and (f) show the difference between two oppositely circularly polarized absorption components of the probe beam. When this signal is non-zero an asymmetry exists in the distribution of angular momentum along the $x$-axis (observation direction). It should be noted here that the green curve is a flat line in (d), (e) and (f) because all ground-state coherences have been eliminated and therefore do not appear in the absorption $\sigma^+ - \sigma^-$ difference signal, even those coherences that originated in the excited state and coherently decayed to the ground state. Potassium shows the highest value of relative absorption difference (Fig.\ref{fig:abs_elements} (d)) out of all three chemical elements shown here. Nevertheless, we deem K atoms not suitable for measurements, because the magnetic field regions where the excited- and ground-state coherences are created are very close to each other in terms of magnetic field, and so the subsequent analysis and interpretation of these signals could prove cumbersome.

The absorption difference signal of $^{85}$Rb atoms (Fig.~\ref{fig:abs_elements}(e)) appears to be less influenced by the excited-state coherent effects when compared to the K (Fig.~\ref{fig:abs_elements}(d)), as can be seen from the small differences between the blue and orange curves (Fig.~\ref{fig:abs_elements}(e)).
But in the range of the magnetic field (1000--2000 G) where the ground-state orientation appears, \textit{i.e.}, where the orange curve coincides with the blue one (Fig.~\ref{fig:abs_elements}(e)), the relative amplitude in terms of the ratio of the absorption difference signal to the total absorption signal of the Rb is smaller when compared to the Cs signal.

The differences between the blue and the orange curves from figure~\ref{fig:abs_elements}(f) show that the absorption difference signals of Cs atoms are influenced by both the excited- and ground-state coherences at low and intermediate magnetic field values, but after approximately 1000~G the signal is almost entirely dependent on ground-state coherences. The fact that the orange and blue curves coincide above 1500 G shows that the contribution of excited-state effects to the blue curve is insignificant.
As a result we conclude that Cs is the most suitable element for observation of angular momentum AOC in the ground state.

Another parameter that the absorption signals depend on is the Rabi frequency of the pump beam. The results from numeric modeling showed that by increasing the Rabi frequency, the absorption signal differences increased in amplitude, but showed little change in shape. All modeled signals in Figure~\ref{fig:abs_elements} have the same Rabi frequency $\Omega_R = 100~\mathrm{MHz}$. Nevertheless, it should be noted that the necessary laser power values will differ slightly between elements because of different dipole transition matrix element values associated with each chemical element~\cite{Auzinsh:2010book} and experimental conditions.

We included the modeled individual circularly polarized absorption components in the upper part of Fig.~\ref{fig:abs_elements} because the interpretation of the origin of the observed signals is more easily understood and to emphasize the scale of the difference between the two absorption components. The amplitudes of the differences of the absorption components are about $10^3$ times smaller than the amplitudes of the individual circularly polarized absorption components.

\subsection{Experimentally observed absorption signals and analysis} \label{sub_B}

\begin{figure}[t]
\includegraphics[width=\linewidth]{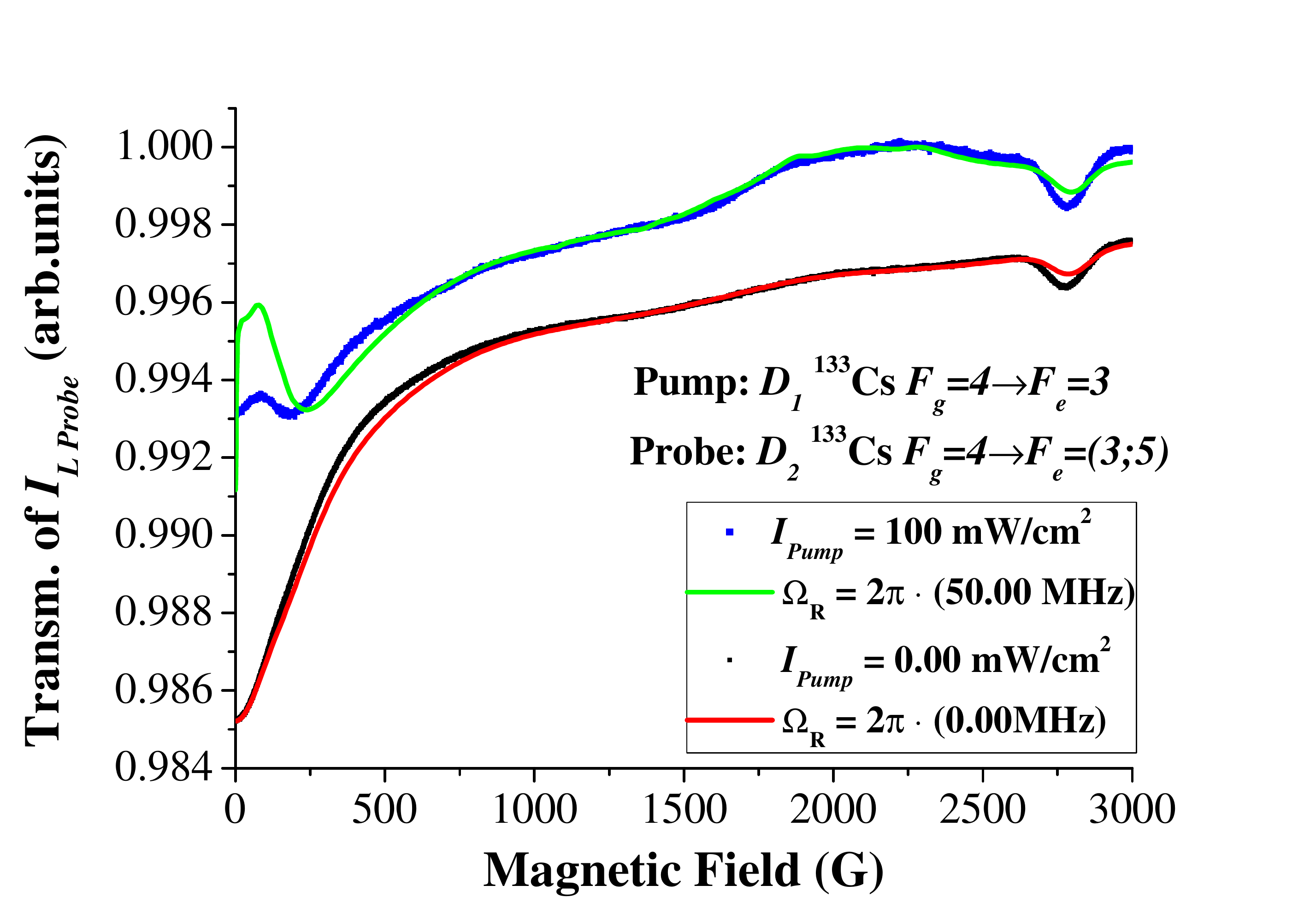}
    \caption{\label{fig:komp}Transmission signal of a single circularly polarized light component ($I_{L Probe}$): Pump laser off: black dots---experimental data; red curve---theoretical calculations; Pump laser on: blue dots---experimental data; green curve---theoretical calculations. $\gamma = 0.019$~MHz}
\end{figure}

Next we show the results of the experimentally measured transmission signal of a single circularly polarized absorption component with the pump beam off and on (Fig.~\ref{fig:komp}). The signal from only one of the two transmission components is shown because the two counter circularly polarized transmission signals are very similar. Figure~\ref{fig:komp} shows the transmission signals when the frequency of the probe beam was fixed to the crossover peak from the SAS signal that corresponds to the $F_g=4 \rightarrow F_e=(3;5)$ transition.

\begin{figure}[h]
	\includegraphics[width=\linewidth]{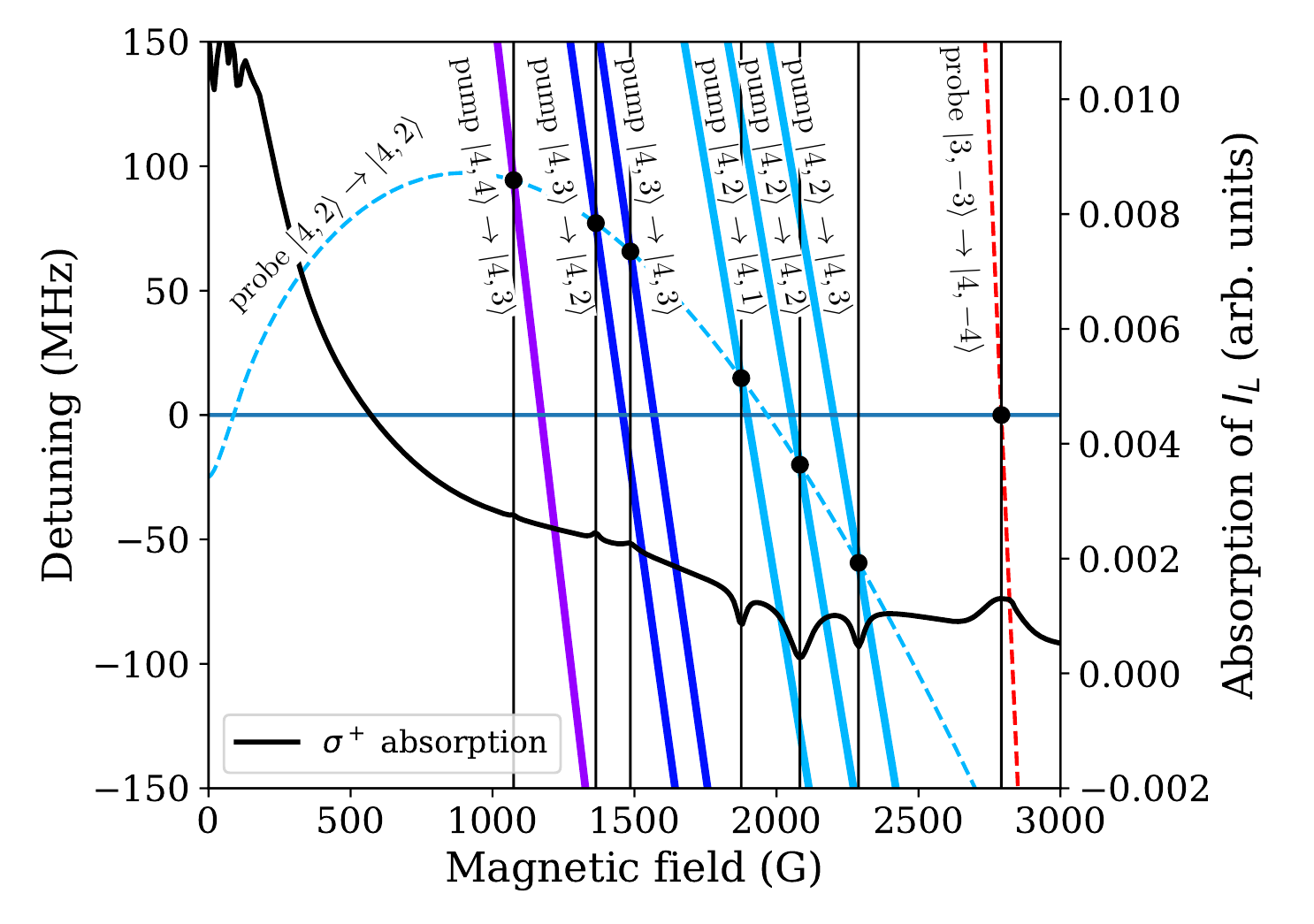}
    \caption{\label{fig:deltani}
    The vertical axis on the right shows the absorption intensity of one of the two circularly polarized components. Black curve (right axis) -- calculated absorption signal of a single circularly polarized light component ($I_{L Probe}$) with the pump laser on: $\Omega_R=5$ MHz.
    The vertical axis on the left shows the detuning of transition frequency between magnetic sublevels from either the pump or probe laser frequency. The colored lines (left axis) represent the dependence of the energy difference between various pairs of magnetic sublevels on magnetic field.}
\end{figure}

Before we describe the physical processes responsible for the shape of transmission signal with the pump beam on, we first describe the transmission signal while the pump beam was switched off.
When the pump beam is off we obtain the experimental transmission signal depicted by the black dots, which is in agreement with the theoretical signal represented by the red curve in Fig.~\ref{fig:komp}. The transmission signal rises as the magnetic field is increased. Initially a rather fast increase is observed, but after around 1000~G the slope of the increase diminishes. While the pump beam is off, notably also when the pump is on, a single pronounced feature can be observed at 2800~G: a peak that shows a decrease in the transmission signal. 
The interpretation of the physical processes responsible for the shapes of the transmission signals from Fig.~\ref{fig:komp} is given in Fig.~\ref{fig:deltani}.
The black curve in Fig.~\ref{fig:deltani} shows the theoretical absorption signal of a single circularly polarized light component with $\Omega_R=5$~MHz and $\gamma=0.019$~MHz in order to emphasize the substructures that underlie the transmission signals in Fig.~\ref{fig:komp}. Note the relative inversion of the absorption and transmission signals. 
The dashed lines in Fig.~\ref{fig:deltani} represent the transition frequencies between pairs of magnetic sublevels from the Cs $D_2$ hyperfine manifold relative to the fixed probe laser frequency, which corresponds to $\Delta\nu=0$. The peak that appears in the observed signal at 2800~G, is caused by two magnetic sub-levels ($m_{F_g=3}=-3 \rightarrow m_{F_e=4}=-4$) from the Cs $D_2$ hyperfine manifold coming into resonance with the probe beam, thus leading to a decrease in the transmission signal (Fig.~\ref{fig:komp} around 2800~G).

Next we switched the pump beam on and observed the change in the transmission signal. The obtained experimental transmission signal is depicted by the blue dots, and the theoretical calculations are the solid green curve in Fig.~\ref{fig:komp}. In this case the frequency of the probe beam was also fixed to the crossover peak of Cs $D_2$ $F_g=4 \rightarrow F_e=(3;5)$ transition, while the frequency of the pump beam was set to the Cs $D_1$ $F_g=4 \rightarrow F_e=3$ hyperfine transition. The overall tendency of the transmission signal to increase stayed the same. Nevertheless, noticeable changes in the transmission signal occur in the region from 0 to 500~G, and another broad structure appears at around 2000~G. The changes in the signal at the low field are strongly influenced by the processes in the excited state. Consequently, no further analysis of the low-magnetic-field changes in the transmission signal will be provided here at this point since we are mostly concerned with the ground-state AOC. The very broad feature that appears at around 2000~G is strongly connected with the ground-state processes. This feature is caused by the interaction of ground-state magnetic sublevels with both lasers. The solid lines in Fig.~\ref{fig:deltani} correspond to the relative transition frequencies between two magnetic sublevels from the Cs $D_1$ hyperfine manifold with respect to the pump laser frequency. When the solid lines of the Cs $D_1$ pump cross the dashed lines of the Cs $D_2$ probe an increase in the transmission signal can be observed (Fig.~\ref{fig:komp} around 2000~G). The transmission increases because the pump beam has emptied the ground-state magnetic sublevel $m_{F_g=4}=2$; thus, the probe beam is not absorbed, and the transmission increases.
Three small peaks that appear as an increase in the simulated absorption signal are present in Fig.~\ref{fig:deltani} (black curve) around 1000--1500~G. The origin of these structures is similar to the features around 2000~G, which we have just described with the exception of the fact that instead of emptying $m_{F_g=4}=2$, in this situation, the pump beam is repopulating this sublevel, because the population is returned to the $m_{F_g=4}=2$ via the $m_{F_e=4}=2$ and $m_{F_e=4}=3$ magnetic sublevels of the Cs $D_1$ manifold with which the pump beam is interacting.

Because the coupling between the total angular momentum of the electron $J$ and the nuclear spin $I$ is almost completely broken up in the excited state of Cs but partially broken in the ground state, the transition probabilities change and "tend toward" the selection rules $\Delta m_J=0,\pm1$ and $\Delta m_I=0$; therefore, Fig.~\ref{fig:deltani} shows only transitions between magnetic sublevels with transition probability that differ from 0. 

\begin{figure}[t]
    \includegraphics[width=\linewidth]{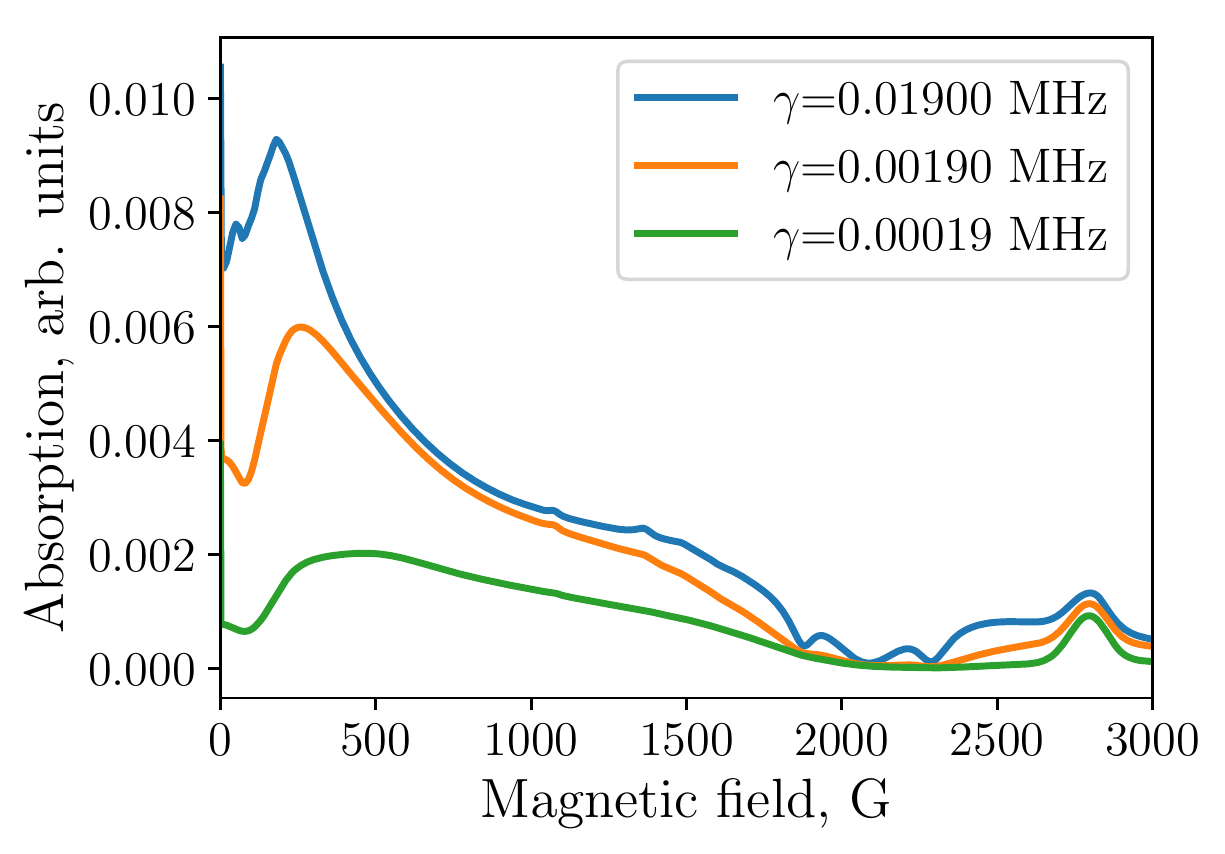}
    \caption{The dependence of the absorption signal of a single circularly polarized light component on the transition relaxation $\gamma$. ($I_{LProbe}$): Pump laser on: $\Omega_R = 20$ MHz.}
    \label{fig:gamma_dependence}
\end{figure}

The corresponding Rabi frequency of the theoretical transmission signal (Fig.~\ref{fig:komp} green solid curve) was estimated from Eq.~\eqref{eq:Rabi} and shows adequate agreement. The black solid curve in Fig.~\ref{fig:deltani} corresponds to the theoretical absorption signal when the Rabi frequency is $\Omega_R=5$~ MHz suggesting that the individual probe-pump resonances around 2000~G could be observed at lower power values of the pump beam. However, the three peaks around 2000~G that appear in the absorption signal (Fig.~\ref{fig:deltani} black curve) could not be reproduced in our experiments.
When estimating the Rabi frequency and comparing the theoretical signals to the ones obtained from the experiment (Fig.~\ref{fig:komp}), we kept the transit relaxation rate $\gamma$ fixed. The estimation of $\gamma$ follows the relationship described in eq.~\ref{eq:gamma}. Because the laser intensity is not constant across the diameter of the pump laser beam, the influence of Rabi frequency and transit relaxation rate is complex \cite{Auzinsh:2016}. Keeping this in mind, we allowed $\gamma$ to vary from the definition in Eq.~\ref{eq:gamma}.
Then, by keeping the Rabi frequency fixed we can observe the influence of $\gamma$ on the observed signal in Fig.~\ref{fig:gamma_dependence}. As the transit relaxation rate is decreased the width of the three peaks around 2000~G increases and the structure becomes unresolved, which is in better qualitative agreement with the experimental signals in Fig.~\ref{fig:komp}.

\subsection{Analysis of the absorption difference signals for imperfect experimental conditions} \label{sub_C}

\begin{figure*}[t]
    \includegraphics[width=0.99\textwidth]{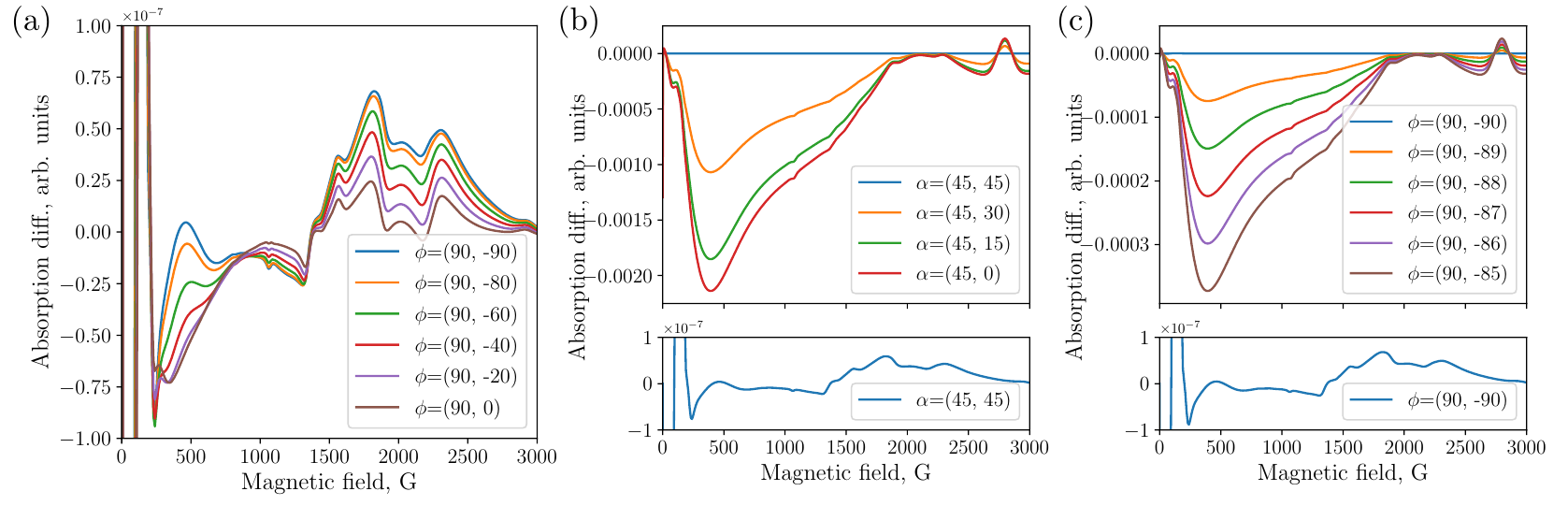}
    \caption{a:~Absorption difference dependence on polarization phase $\phi$ ($\alpha=[45^{\circ}, 45^{\circ}]$). b:~Absorption difference dependence on polarization angle $\alpha$. Bottom figure shows absorption difference for $\alpha=[45^{\circ}, 45^{\circ}]$ in a zoomed in scale ($\phi=[60^{\circ}, -60^{\circ}]$). c:~Absorption difference dependence on polarization phase $\phi$, if the second component is a vertical ellipse with $\alpha=0^\circ$. Bottom figure shows absorption difference for $\phi=90^\circ$ and  $\phi=-90^\circ$ in a zoomed in scale ($\alpha=[45^{\circ}, 0^{\circ}]$). $\Omega_R=20$~MHz for all signals.}
    \label{fig:pol_dependence}
\end{figure*}

\begin{figure}[t]
    \includegraphics[width=0.99\linewidth]{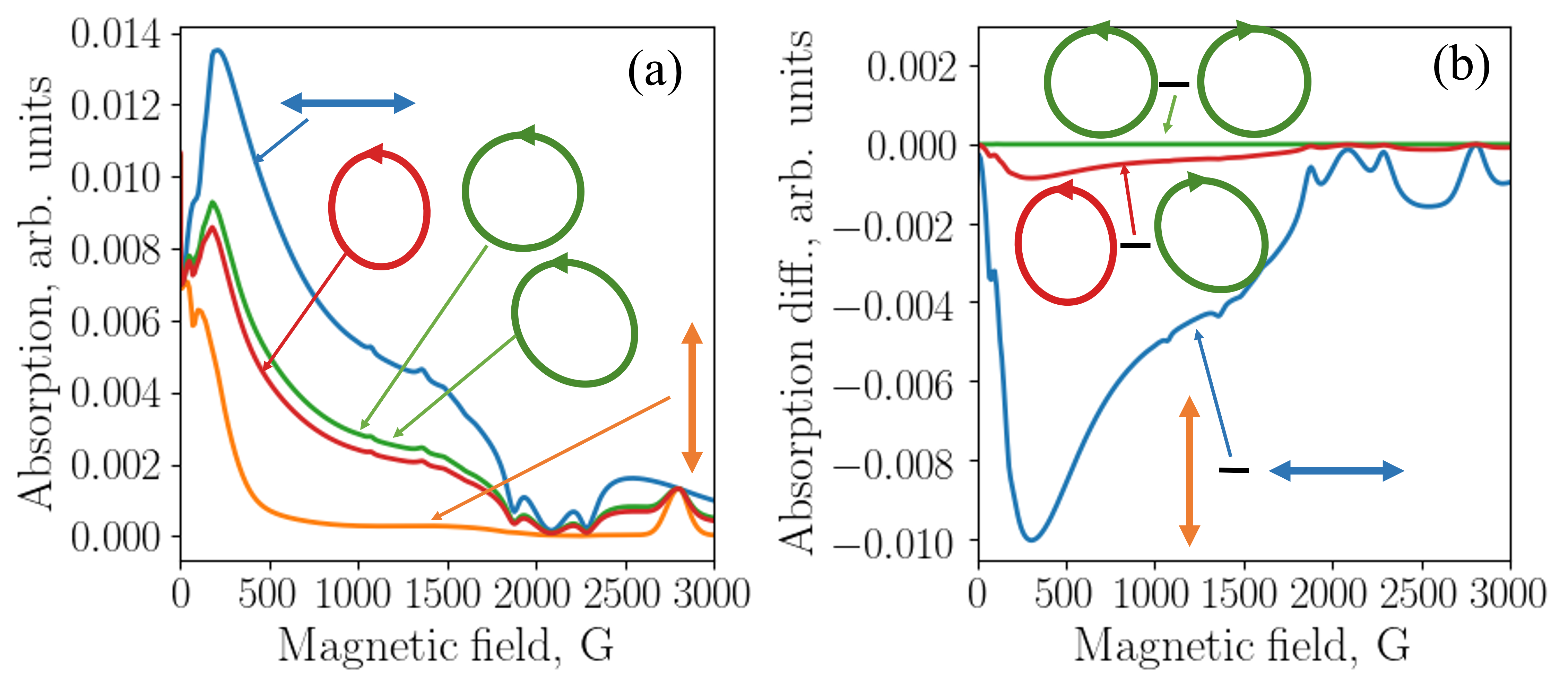}
    \caption{a:~Numerical calculations of individual absorption components with different probe beam polarizations.
    b:~Difference signals of individual absorption components with different probe beam polarizations.
    The polarization of each singal is indicated in the figure.
    }
    \label{fig:ellipse}
\end{figure}

Experimentally it is very difficult to obtain $100\%$ circularly polarized light. Therefore an analysis of the theoretical results of the difference signal of the two oppositely circularly polarized absorption components was carried out with respect to the polarization properties of the probe beam, keeping the pump beam unchanged. The Rabi frequency of the pump beam was set to 20~MHz and $\gamma$ was set to 0.019~MHz for all calculations in the subsequent analysis, results of which are shown in Fig.~\ref{fig:pol_dependence}. To obtain the different polarization states of the probe beam we use two variables: the phase between the two constituent linearly polarized components $\phi=\phi_z - \phi_y$, and the azimuth angle $\alpha$, which is the angle between the major axis of the polarization ellipse and the quantization axis $z$ in the $zy$ plane (see inset of Fig.~\ref{fig:geom}).

First, we inspected how the difference signal depends on the phase $\phi$.  
Figure~\ref{fig:pol_dependence}(a) shows the results from theoretical calculations of the difference signals for the case when one of the nominally circularly polarized light components was assumed to be completely circularly polarized with a fixed phase value of $\phi = 90^{\circ}$ while the phase of the other component was changed from $-90^{\circ}$ to $0^{\circ}$. As a result the polarization state of the other component changed from completely circularly polarized to elliptically polarized with a fixed azimuth angle $\alpha = 45^{\circ}$, and finally to linearly polarized. The latter case of phase $\phi = 0^{\circ}$ corresponds to linearly polarized light at an azimuth angle of $\alpha = 45^{\circ}$ with respect to the quantization axis in the $zy$ plane.
In Fig. 9(a), varying the phase difference $\phi$ over the entire range causes changes less than one part in $10^{-7}$, whereas changes in the azimuth angle $\alpha$ (Fig. 9 (b)) or $\phi$ with a fixed $\alpha=45$\degree  for one component (Fig. 9 (c)) cause changes that are 4 orders of magnitude larger.

Second, we determined the impact of the azimuth angle $\alpha$ on the difference signal. In order to observe the impact of the azimuth angle more clearly, we assumed that the polarization state of both probe beam components to be imperfect by setting the phase values to $60^{\circ}$ and $-60^{\circ}$, respectively. Then we changed the azimuth angle $\alpha$ of one of the components from $0^{\circ}$ to $-45^{\circ}$, while keeping the azimuth angle of the other component fixed to a constant value of $45^{\circ}$ with respect to the quantization axis in the $zy$ plane, as shown in the inset of Fig.~\ref{fig:geom}. Figure~\ref{fig:pol_dependence}(b) (top) clearly shows the difference signal to be strongly dependent on the azimuth angle. A change in the azimuth angle of $15^{\circ}$ (orange curve Fig.~\ref{fig:pol_dependence}(b)) changes the amplitude of the signal by 4 orders of magnitude. Note the different vertical scales of figure~\ref{fig:pol_dependence}(b) top and bottom. Furthermore, the shape of the difference signal changes from the shape that was obtained from two oppositely circularly polarized light components to a shape that resembles the difference between two orthogonal linearly polarized light components with azimuth angles $\pm45^{\circ}$, \textit{i.e.}, the blue curve in Fig.~\ref{fig:ellipse}(b).
When the azimuth angles of the elliptically polarized components are not aligned, the difference of linear components starts to dominate.

Third we analyzed a more realistic case when one of the components is circularly polarized, but both the phase and azimuth angle of the other component are set to values that differ from the ideal values for circularly polarized light. Here we assumed that the first component was circularly polarized with a fixed phase value $90^{\circ}$.
As we decreased the value of $\phi$ for the second component we kept the azimuth angle $\alpha=0^{\circ}$ fixed (vertical).
Figure~\ref{fig:pol_dependence}(c) shows that by changing the phase of the second component as little as $1^{\circ}$ the amplitude of the difference signal increases by 3 orders of magnitude (note the different vertical scales of figure~\ref{fig:pol_dependence}(c) top and bottom).
In reality it is likely that both components are with imperfect azimuth angle and phase, which could degrade the difference signal from the ideal case even more.

Figure~\ref{fig:ellipse}(a) shows the relative amplitudes of the linear absorption components (blue and orange curves) of the probe beam propagating along the $x$-axis in comparison to the individual circularly polarized absorption components of the probe beam propagating in the same direction. Figure~\ref{fig:ellipse}(b) shows the difference between the two orthogonal linearly polarized absorption components (blue curve) relative to the difference of the circularly polarized absorption components. The dominance of the difference of the linearly polarized components can be clearly seen in Fig.~\ref{fig:ellipse}. The larger amplitude of the difference of the linearly polarized components means that while ground-state AOC is created, simultaneously also a ground-state angular momentum alignment exists in the $zy$ plane, but the ground-state orientation is 5 orders of magnitude smaller.

\section{Conclusions}

We have modeled absorption signals for different alkali atoms and by separating the ground and excited state coherent effects we determined that the ground-state AOC could be most easily observed in Cs atoms because the absorption signals show little influence of the excited state.
We have interpreted the shapes of these signals in terms of the resonance conditions between magnetic sublevels and frequencies of two lasers.
We have studied how these signals depend on several experimental parameters, e.g., the pump laser intensity and diameter, and showed that the absorption signal is extremely sensitive to the probe light polarization.

By sequentially probing the system with two counter-circularly polarized laser beams, we have experimentally observed the ground-state angular momentum distribution along the probe beam, which was created by a strong linearly polarized pump laser and an external magnetic field in the ground state of Cs, and these experimental results are in good agreement with theoretical signals. In addition, we have provided an explanation for the physical processes responsible for the shapes of these absorption signals.

The absorption difference ($\sigma^+-\sigma^-$) signal allows for direct detection of orientation in the ground state and a straightforward interpretation of the signals, but the preparation of the properties of the probe beam is challenging. Another way to detect transverse orientation is to observe fluorescence which contains only circularly polarized light when transitions to both ground-state hyperfine levels are observed \cite{Auzinsh:2009b}. Although the fluorescence signal is not contaminated with linearly polarized light, the interpretation of the signal is not as straightforward as for the absorption signal, and so a slightly more complex analysis is necessary \cite{Mozers:2020}.

Due to the fact that the difference of the circularly polarized absorption components is strongly dependent on the polarization properties of the probe beam, actually it is impossible to measure this signal without extremely high-precision control over the polarization of the probe beam. Therefore, we deem the difference of the circularly polarized fluorescence signals more suitable for the detection of ground-state AOC despite the necessary complex analysis.
However, there exists another way to detect the AOC in the ground state: if a spectrally broad light source is used for the absorption signals, wide enough to cover all of the excited-state hyperfine levels, then the difference signal will contain only the information about the angular momentum orientation because the spectrally broad probing of the system can sense multipoles only as high as $\Delta m_F=\pm1$.

Besides the AOC effects considered in this paper, the appearance of the left-right asymmetry in atoms which leads to a circularly polarized fluorescence or differences in absorption of two opposite circularly polarized probe beams can be associated with parity violation (PV) related to the weak interaction (see T. D. Lee and C. N. Yang \cite{Lee:1956}). Parity violation associated with the weak interaction manifests itself most prominently in the $\beta$-decay of polarized $^{60}$Co nuclei \cite{Wu:1957}, but can be observed in atoms as well \cite{Roberts:2015}.

Cs atoms often are used for atomic PV measurements as well, see for example \cite{Bouchiat:2012}. In the cited experiment atoms were subjected simultaneously to the orthogonal electric and magnetic fields. Additionally to the stationary fields, in these PV experiments atoms simultaneously interact with two linearly polarized laser fields, one of which is tuned to the vicinity of the first Cs resonant D$_1$ $6{^2}S_{1/2} \rightarrow 6^{2}P_{1/2}$ transition and second to the dipole forbidden transition. The present study shows that if the laser fields and external magnetic field are not exactly orthogonal, AOC can appear, and the resultant effect---appearance of right-left asymmetry---can be rather similar to effects caused by PV. Interpretation of experimental results of PV measurements still continues~\cite{Sahoo:2021}. This gives additional motivation to study AOC effects in detail in order to prevent misinterpretation of the observed effects.

Another class of experiments that is vulnerable to an AOC background relates to the search for a permanent electric dipole moment (EDM) of an electron. Similar to the experiments aimed at PV detection, in EDM experiments~\cite{Rochester2001, khriplovich1997cp} performed with atoms and polar diatomic molecules~\cite{Cairncross2019} even a very small misalignment between applied external magnetic and electric fields can complicate the interpretation of the results due to AOC effects. Therefore, a detailed understanding of AOC is helpful for designing and interpreting results from such experiments correctly.

\begin{acknowledgments}
A.M. acknowledges support from PostDoc Latvia Grant No. 1.1.1.2/16/117 “Experimental and theoretical signals of ground-state angular momentum alignment-to-orientation conversion by the influence of laser radiation and external magnetic field in atomic alkali metal vapour.” L.B., D.O., F.G. and M.A. acknowledges support from Latvian Council of Science project “Compact 3-D magnetometry in Cs atomic vapor at room temperature” Project No. lzp-2020/1-0180.
\end{acknowledgments}

%\appendix
%\section{Appendixes}

% \nocite{*}

\bibliography{main_AOC_Cs_Abs}% Produces the bibliography via BibTeX.

\end{document}